\documentstyle[NATO,psfig,namedreferences]{crckapb} 

\newcommand{\stt}{\small\tt}

\def\lsim{\mathrel{\mathpalette\vereq<}}
\def\gsim{\mathrel{\mathpalette\vereq>}}
\catcode`\@=11
\def\vereq#1#2{\lower3pt\vbox{\baselineskip1.5pt \lineskip1.5pt
\ialign{$\m@th#1\hfill##\hfil$\crcr#2\crcr\sim\crcr}}}
\catcode`\@=12
\begin{opening}
\title{THE GLOBAL COSMOLOGICAL PARAMETERS}

\author{MASATAKA FUKUGITA}
\institute{University of Tokyo, Institute for Cosmic Ray Research\\ 
Tanashi, Tokyo 188, Japan, and\\
           Institute for Advanced Study, Princeton, NJ 08540, U. S. A.}

\end{opening}

\runningtitle{THE CRCKAPB STYLE FILE}

\begin{document}
\input epsf 


\section{Introduction}

In these lectures I shall discuss the status of the determination
of the three cosmological parameters 
which enter the Einstein equation and govern geometry
and evolution of space-time of the Universe: 
the Hubble constant $H_0$,
the mass density parameter $\Omega$ and the cosmological constant
$\lambda$. 

Among the three parameters,
the Hubble constant is the dimensionfull quantity which sets the
basic size and age of the Universe. The perennial effort
to determine $H_0$ dates back to Hubble (1925) and has 
a long history of disconcordance.  Recent progress has
done much to resolve the long-standing
discrepancy concerning the extragalactic distance scale, but there are 
some newly revealed uncertainties in the distance scale
within the Milky Way. The emphasis in this lecture is on discussion of 
these uncertainties. 

The mass density parameter directly determines the formation of
cosmic structure. So, as our understanding of the cosmic structure 
formation is tightened, we should have a convergence of the
$\Omega$ parameter. An important test is to examine whether the  
$\Omega$ parameter extracted from cosmic structure formation 
agrees with the value estimated in more direct ways. 
This gives an essential verification for the theory of structure formation.

The third important parameter in the Friedmann universe is the cosmological
constant $\Lambda$. We now have some evidence for a 
non-zero $\Lambda$ which, if confirmed, would have
most profound implications for 
fundamental physics. This lecture will focus on 
the strength of this `evidence'.

We take the normalisation

\begin{equation}
\Omega+\lambda=1
\end{equation}
for the flat curvature, where $\lambda=\Lambda/3H^2_0$ with  
$\Lambda$ the constant entering in the Einstein equation.
The case with $\Omega=1$ and $\lambda=0$ is referred to as 
the Einstein-de Sitter (EdS) universe. 
We often use distance modulus 

\begin{equation}
m-M=5\log (d_L/10{\rm pc})
\end{equation}
instead of the distance $d_L$. For conciseness, we shall omit the units 
for the Hubble constant, (km s$^{-1}$Mpc$^{-1}$).

After the Summer Institute there appeared several important papers on
the distance scale. I try to incorporate these results in this article. 

\section{The Hubble Constant}

\subsection{Historical note}

The global value of $H_0$ has long been uncertain by a factor of
two. Before 1980 the dispute was basically between two schools: 
Sandage and collaborators had insisted on $H_0=50$ (Sandage \& Tammann
1982); de Vaucouleurs and collaborators preferred a high value $H_0=90-100$ 
(de Vaucouleurs 1981). Conspicuous progress was brought by the
discovery of an empirical but tight relationship between galaxy's luminosity 
and rotation velocity,
known as the Tully-Fisher relation (Tully \& Fisher 1977). The use
of the Tully-Fisher relation has largely reduced subjective
elements in the distance work, and $H_0=80-90$ has been derived
from a straightforward reading of the Tully-Fisher 
relation.  Representative of this work are the papers 
of Aaronson et al.
(1986) and Pierce \& Tully (1988). A doubt was whether the result was
marred with the Malmquist bias --- whether the sample selects 
preferentially bright galaxies, and hence the result was biased towards
a shorter distance (Kraan-Korteweg, Cameron \& Tammann 1988; Sandage 1993a). 
A related dispute was over the distance to the Virgo
cluster, whether it is 16 Mpc or 22 Mpc: 
the different results depending on which
sample one used.

The next momentous advancement was seen in 1989$-$1990 when a few 
qualified distance indicators were discovered. One of them is a
technique using planetary nebula luminosity function (PNLF), the shape of
which looked universal (Jacoby et al. 1990a). 
Another important technique is the use of
surface brightness fluctuations (SBF), utilizing the fact that the images of
distant galaxies show a smoother light distribution; while surface brightness
does not depend on the distance, pixel-to-pixel fluctuations in a  
CCD camera decreases as $d_L^{-1}$
(Tonry \& Schneider 1988). They proposed that this smoothness can 
be a distance indicator if the stellar population is uniform. 
What was important is that the two completely independent
methods predicted distances to individual galaxies in excellent agreement
with each other (Ciardullo, Jacoby \& Tonry 1993). 
The PNLF/SBF distance also agreed with the value from the
Tully-Fisher relation, with a somewhat larger scatter. 
These new techniques,
when calibrated with the distance to M31, yielded a value around $H_0=80$
and the Virgo distance of 15 Mpc
(For a review of the methods, see Jacoby et al. 1992).

Around the same time the use of Type Ia supernovae (SNeIa) became popular
(Tammann \& Leibundgut 1990; Leibundgut \& Pinto 1992; Branch \& Miller 1993).
The principle is that the maximum brightness of SNIa is nearly 
constant, which can be used as an absolute standard candle.
Arnett, Branch and Wheeler proposed that
the maximum brightness is reliably calculable using models which are
constrained from observations of released kinetic energy
(Arnett, Branch \& Wheeler 1985;
Branch 1992). This led to  $H_0=50-55$, in agreement with the 
calibration based on the first Cepheid measurement of the nearest 
SNIa host galaxy using the pre-refurbished 
{\it Hubble Space Telescope} (HST) (Sandage et al.
1992).
In the early nineties the discrepancy was dichotomous as whether $H_0=80$
{\it or} 50. (see Fukugita, Hogan \& Peebles 1993 for the status at that
time; see also van den Bergh 1989, 1994).
      
The next major advancement was brought with the refurbishment mission 
of HST, which enabled one to resolve
Cepheids in galaxies as distant as 20 Mpc (1994). This secured the
distance to the Virgo cluster and tightened the 
calibrations of the extragalactic distance indicators,
resulting in $H_0=(70-75)\pm10$, 10\% lower than the `high value'.
Another important contribution was the discovery that the 
maximum brightness of SNeIa  varies
from supernova to supernova, and that it correlates with the decline rate
of brightness (Pskovski$\check\i$ 1984; 
Phillips 1993; Riess, Press \& Kirshner 1995;
Hamuy et al. 1996a). 
This correction, combined with the direct calibration of the maximum brightness
of several SNeIa with HST Cepheid observations, raised the `low value' of
$H_0$ to $65 {+5 \atop -10}$, appreciably higher than 55.
This seemed to resolve the long-standing controversy. 

All methods mentioned above use distance ladders and take the 
distance to Large Magellanic Clouds (LMC) to be 50 kpc ($m-M=18.5$)
as the zero point. Before 1997 few doubts were cast on the
distance to LMC (TABLE 1 shows a summary of the distance to LMC known as
of 1997). With the exception of RR Lyr, the distance converged to
$m-M=18.5\pm0.1$, i.e., within 5\% error, and the discrepency of the 
RR Lyr distance was blamed on
its larger calibration error.
It had been believed that the Hipparcos mission (ESA 1997) would secure the
distance within MW and tighten the distance to
LMC. To our surprise, the work using the Hipparcos 
catalogue revealed the contrary; the distance to LMC was
more uncertain than we had thought, introducing new difficulties 
into the determination of $H_0$. 
In this connection, the
age of the Universe turned out to be more uncertain than it was believed.

During the nineties, efforts have also been conducted
to determine the Hubble constant without
resorting to astronomical ladders. They are called `physical methods'.
The advantage of the ladder is that the error of each ladder can be documented
relatively easily, while the disadvantage is that these errors accumulate.
Physical methods are free from the accumulation of errors, but on the other
hand it is not easy to document the systematic errors. Therefore, the
central problem is how to minimise the model dependence and
document realistic systematic errors. Nearly ten years of effort
has brought results that can be compared with the distances 
from ladders. The physical methods include the expansion photosphere 
model (EPM) for type II SNe 
(Schmidt, Kirshner \& Eastman 1992) and gravitational lensing time
delay (Refsdal 1964). Use of SNeIa maximum brightness was once taken to be a
physical method (Branch 1992), but then `degraded' to be a ladder, which 
however significantly enhanced its accuracy.   

\begin{table}[htb]
\begin{center}
\caption{Distance to LMC as of 1997}
\begin{tabular}{lll}
\hline
Method & Ref & Distance moduli \\
\hline
Cepheid optical PL & Feast \& Walker 1987 & 18.47$\pm$0.15\\
Cepheid optical PL & Madore \& Freedman 1991 & 18.50$\pm$0.10\\
Cepheid IR PL      & Laney \& Stobie 1994 & 18.53$\pm$0.04\\
Mira PL            & Feast \& Walker 1987 & 18.48$\pm$(0.06)\\
SN1987A ring echo  & Panagia et al. 1991   & 18.50$\pm$0.13\\
SN1987A EPM        & Schmidt et al. 1992   & 18.45$\pm$0.13\\
RR Lyrae           & van den Bergh  1995   & 18.23$\pm$0.04\\
\hline
\end{tabular}
\end{center}
\end{table}

\subsection{Extragalactic distance scale}

The measurement of cosmological distances traditionally employs
distance ladders (see Weinberg 1972). The most 
traditional ladders are shown in TABLE 2. 
The listings written in italic indicate new
methods which circumvent intermediate rungs. 
The most important milestone of the ladder is LMC at 50kpc ($m-M=18.5)$.  
A distance indicator of particular historical importance 
(Hubble 1925)
is the Cepheid period-luminosity (PL) relation, which is given 
a great confidence, but
we note that it requires 
a few rungs of ladders to calibrate its zero 
point. 

\begin{table}[htb]
\begin{center}
\caption{Traditional distance ladders}
\begin{tabular}{lll}
\hline
Method & Distance range & typical targets\\
\hline
{\bf Population I stars} & & \\
trigonometric or kinematic methods (ground)    & $<$50 pc & Hyades, nearby dwarfs\\
main sequence fitting (FG stars) Pop. I   & $<$200pc & Pleiades \\
{\it trigonometric method (Hipparcos)}   &  $<$500pc & nearby open clusters\\
main sequence fitting (B stars) & 40pc$-$10kpc & open clusters\\
Cepheids [Population I] (ground) & 1kpc$-$3Mpc & LMC, M31, M81\\
{\it Cepheids} [{\it Population I}]{\it (HST)} & $<$30Mpc & Virgo included\\
secondary (extragalactic) indicators & 700kpc$-$100Mpc\\
\hline
{\bf Population II stars} & & \\
{\it trigonometric method (Hipparcos)}   &  $<$500pc & nearby subdwarfs\\
subdwarf main sequence fitting &100pc$-$10kpc& global clusters\\
cluster RR Lyr & 5kpc$-$100kpc& LMC, age determinations\\ 
\hline
\end{tabular}
\end{center}
\end{table}

Prior to the HST work there were only 4$-$5 galaxies with Cepheid
distances which could be used to calibrate secondary indicators. 
The reach of the ground-based Cepheid measurement is about 3 Mpc, which
means that one cannot increase the number of calibrating galaxies from
the ground. Pierce et al. (1994) could finally measure Cepheids in 
NGC 4571 in the Virgo cluster at 15 Mpc, but only with the best seeing 
conditions and difficult observations.   
The refurbishment of HST achieved a sufficient power
to resolve Cepheids at the Virgo cluster (Freedman et al. 1994). 
Now 28 nearby spiral galaxies within 25 Mpc are given 
distances measured using the Cepheid PL relation (Ferrarese et al. 1999b).
A typical random error is 4-5\% (0.08-0.10 mag), and the systematic
error (from photometry) 
is 5\% (0.1 mag) excluding the uncertainty of the LMC distance,
to which the HST-Key Project(KP) group assigns 6.5\% error (0.13 mag). 
The prime use of these galaxies is to calibrate secondary distance
indicators which penetrate into a sufficient depth that perturbations in
the Hubble flow are small enough compared with the flow itself.

Cepheids are Population I stars, so reside only in spiral galaxies.
The calibration is therefore direct for TF and some SNeIa. For early
type galaxies (fundamental plane or $D_n-\sigma$, and SBF) the 
calibration is not very tight;
one must either use some groups where both
early and late galaxies coexist, or regard the bulges of spiral galaxies
as belonging to the same class as early galaxies and avoid contaminations
from discs. 
Additional observations have been made for the galaxies that host
SNeIa (Saha et al. 1999).
The results are summarised in TABLE 3. We include a few earlier SNIa
results which employ a partial list of Cepheid calibrators.

\begin{table}[htb]
\begin{center}
\caption{Hubble constant}
\begin{tabular}{lll}
\hline
Secondary indicators & Refs & Hubble constant \\
\hline
Tully-Fisher & HST-KP (Sakai et al. 1999) & 71$\pm 4\pm7$ \\
Fundamental Plane & HST-KP (Kelson et al. 1999) & 78$\pm 8\pm10$ \\
SBF & HST-KP (Ferrarese et al. 1999a) & 69$\pm 4\pm6$ \\
SBF & Tonry et al. (1999) & \underline{77$\pm 4\pm7$} \\
SNeIa & Riess et al. (1995) & 67$\pm$7 \\
SNeIa & Hamuy et al. (1996b) & 63$\pm 3\pm3$ \\
SNeIa & Jha et al. (1999) & 64.4${+5.6 \atop -5.1}$\\
SNeIa & Suntzeff et al. (1999) & 65.6$\pm$1.8 \\
SNeIa & HST-KP (Gibson et al. 1999) & \underline {68$\pm 2\pm5$} \\
SNeIa & Saha et al. (1999)  &60$\pm2$  \\
\hline
Summary (see text)&     &  $(64-78)\pm7$  \\
\hline
\end{tabular}
\end{center}
\end{table}

We accentuate the results with the two methods, 
SBF and SNeIa, in particular to those we underlined in the
table.  
A cross correlation analysis showed that the relative 
distances agree well between SBF and others, including the Cepheid
(Tonry et al. 1997; Freedman et al. 1997),
and that it is probably the best secondary indicator presently
available together with SNeIa; Also important is that 
there are now 300 galaxies measured with SBF, which are essential to
make corrections for peculiar velocity flows for their $\leq4000$
km s$^{-1}$ sample (Tonry et al. 1999).
(PNLF is an indicator of
comparable quality, but it requires more expensive observations so that
applications are rather limited; see Jacoby et al. 1996 for the recent work.)
The final value of Tonry et al. from their $I$ band survey 
is $H_0=77\pm8$, in which $\pm4$
is allotted to uncertainties in the flow model and another $\pm4$
to SBF calibration procedure in addition to the error of the Cepheid
distance $\pm6$ (a quadrature sum is taken).
There are a several other pieces of the SBF work to determine $H_0$, 
which generally
result in $H_0=70-90$ (e.g., Thomsen et al. 1997 using
HST; Jensen et al. 1999 with $K$ band; 
see a review by Blakeslee et al. 1998). 
The new calibration made by the HST-KP group (Ferrarese et al. 1999a)
would decrease $H_0$ only by
2\%. The difference in the final $H_0$ between Tonry et al. and 
Ferrarese et al. comes from using different targets (the latter authors use
only 4 clusters) and flow models.

It is impressive that analyses of SNeIa Hubble diagram
give virtually the same answer, even though the samples are all 
derived primarily from the Cal\'an-Tololo sample of Hamuy et al. (1996b).  
A smaller $H_0$ of Saha et al. (1997) basically reflects the absence of the 
the luminosity-decline rate correction, which pushes up $H_0$ by 10\%.
The other notable difference is a slightly higher value of 
HST-KP (Gibson et al. 1999), who made a reanalysis for all Cepheid
observations performed by other groups and showed that their 
distances (to SN host galaxies) are
all farther than would be derived from the HST-KP procedure. The average
offset is as large as 0.16 mag (8\%). This correction applies
to all results other than HST-KP should we keep uniformity
of the Cepheid data reduction. This is important especially when
one compares the SN results with those from other 
secondary distance indicators, since the calibrations for the
latter exclusively use HST-KP photometry.
Taking the luminosity-decline rate correlation to be real and adopting 
Cepheid distance from the HST-KP data reduction, I adopt $H_0=68$ from SNeIa. 

We present two plots in Figure 1,
(a) the estimates of maximum brightness of different authors and 
(b) the decline rate 
$\Delta m_{15}$, the amount of the decrease in brightness over 
15 days following maximum light, 
both as a function of metallicity [O/H]. The second plot shows how  
metallicity effects are absorbed by the $M_V^{\rm max}-\Delta m_{15}$
relation and the first proves that there is little metallicity dependence
in the corrected maximum brightness, though some scatter is seen among authors.

Leaving out the uncertainty of the Cepheid distance, $H_0$ from
Tonry et al.'s SBF is 77$\pm$6, and that from SNeIa (HST-KP) is
68$\pm$4. The difference is 13\%, and the two values overlap at
$H_0=71$. Allowing for individual two sigma errors, the overlap is in
a range of $H_0=65-76$. An 
additional uncertainty is 6\% error ($\delta H_0=\pm4.5$)
from the Cepheid distance which is common to both, still excluding the
uncertainty of the LMC distance. We may summarise $H_0=71\pm7$ or 64$-$78
as our current standard, provided that LMC is at 50 kpc. 
All numbers in the table are within this range, 
except for the central value of Saha et al (1997). 

In passing, let us note that $H_0=75\pm15$ (Freedman et al. 1997)
obtained directly from the Cepheid
galaxy sample agrees with the global value, implying that peculiar
velocities are not so large even in a 10$-$20 Mpc region.

This convergence is a great achievement, but keep in mind 
that the SNeIa results are still lower than  
those from other secondary indicators\footnote[1]{A remark is given to 
the TF distance. While Sakai et al. (HST-KP) derived
$H_0=71\pm8$ using Giovanelli et al.'s (1997) 
cluster sample, Tammann and collaborators (Tammann 1999; Sandage \& Tammann
1997)  
insist on a low value $H_0=53-56$.
Their cluster result (Federspiel et al. 1998) neglects  the 
depth effect of the Virgo cluster: contrary to ellipticals,
spiral galaxies are distributed 
elongating along the line of sight (Yasuda et al. 1996). Hence 
identifying
the centre of gravity of the spiral galaxy distribution 
with the true core leads to an offset. In fact the presence of
substructure behind the Virgo core is confirmed with the
Cepheid for NGC4649. 
Tammann et al.'s field result 
comes from the allocation of an unusually large dispersion
to the TF relation,
which largely amplifies the Malmquist bias. Tully (1999) obtained $H_0=82\pm16$
(Tully et al. 1998).} by 10\%.
There are additional problems.
First, all these analyses are based on a LMC distance modulus of
$m-M=18.50$ (Feast \& Walker 1987; Madore \& Freedman 1991), 
which has recently been cast into doubt. 
In addition, metallicity effects could lead to systematic errors.
Finally, should we derive the Hubble
constant with the error of 10\%, the problem of dust 
extinction could be an issue, and it is 
potentially coupled with the metallicity. 
We now consider these issues in greater detail.

\begin{figure}
\begin{center}
\centerline{{\epsfxsize=10cm\epsfbox{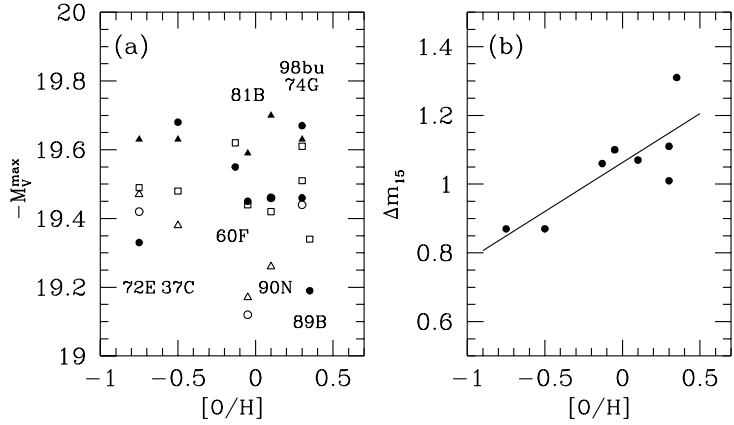}}}
\vspace{-3.7cm}
\caption{(a) Maximum brightness of SNeIa (in the $V$ band)
adopted by different authors
(see TABLE 3) as a function of [O/H] of host galaxies: 
solid circles, Gibson et al. (1999);
solid triangles, Suntzeff et al. 1999; open circles, Jha et al. (1999);
open triangles, Hamuy et al. (1996b); open square, Saha et al. (1999). 
Note that Saha et al.'s calibration is not necessarily brighter, which is
mainly due to a different treatment of extiction.
(b) The decline rate $\Delta m_{15}$ measured in the $B$ band
(Phillips et al. 1999) as a function of [O/H] 
(Gibson et al. 1999). The slope of the curve is 
$\partial \Delta m_{15}/\partial {\rm [O/H]}\simeq 0.28$.
}
\end{center}
\vspace{-1cm}
\end{figure}

\subsection{Distance to LMC}

The present status of the LMC distance is given in TABLE 4.
The most traditional paths to the LMC distance follow the
ladder shown in the upper half of TABLE 2. The Hipparcos
satellite can measure a parallax down to 2 milli arcsec (mas),
corresponding to a distance of 500 pc (ESA 1997). It was a reasonable
expectation that one could obtain the geometric distance to the 
Pleiades cluster, circumventing the main sequence fitting from
nearby parallax stars to the Pleiades and thus securing 
the Galactic distance scale. Hipparcos observations have
also opened a number of novel methods that can be used to
estimate the distance to LMC. This and related activities,
however, 
have actually brought confusions, rather than securing the
distance scale within the MW. We discuss several issues in
order.

\begin{table}[htb]
\begin{center}
\caption{Distance to LMC: Year 1997/1998}
\begin{tabular}{lll}
\hline
Method & Ref & Distance moduli \\
\hline

Cepheid PL & Feast \& Catchpole 1997 &    18.70 $\pm$ 0.10 \\
           & Paturel et al.     1997        &    18.7   \\
           & Madore \& Freedman 1998 &    18.57 $\pm$ 0.11 \\
           & Luri et al. 1998        &    18.29 $\pm$ 0.17 \\
           & Luri et al. 1998        &    18.21 $\pm$ 0.20 \\
           & (traditional) w/ new Pleiades & 18.26 \\
RR Lyrae (stat. para)  & Fernley et al. 1998  &  18.31 $\pm$ 0.10 \\
           &  Luri et al.    1998  &      18.37 $\pm$ 0.23 \\
           & Udalski 1998/Gould et al. 1998 &   18.09 $\pm$ 0.16 \\
RR Lyrae (subdwarf)  & Reid 1997  &       18.65 \\
           & Gratton et al. 1997  &       18.60 $\pm$ 0.07 \\
Mira &  van Leeuwen et al. 1997 &   18.54  $\pm$ 0.18 \\
           & Whitelock et al.   1997  &   18.60 $\pm$  0.18 \\
Red clump  & Udalski et al. 1998a  &       18.08 $\pm$ 0.15 \\
           & Stanek et al. 1998   &      18.07 $\pm$ 0.04\\
           & Cole 1998            &	 18.36 $\pm$ 0.17\\
Eclipsing binaries & Guinan et al. 1998 & 18.30 $\pm$ 0.07\\
           & (Udalski et al. 1998b) &      18.19 $\pm$ 0.13(?)\\
SN 1987A Ring echo & Gould \& Uza 1998   &  $<$18.37 $\pm$ 0.04\\
           &  Sonneborn et al. 1997   &   18.43 $\pm$ 0.10 \\
           & Panagia et al. 1997    &     18.58 $\pm$ 0.03 \\
           & Lundqvist \& Sonneborn1997 & 18.67 $\pm$ 0.08 \\
Cepheid PL & Sekiguchi \& Fukugita 1998 & 18.10-18.60 \\
           & Sandage et al. 1999        & 18.57$\pm$0.05 \\
Cepheid PL (BW method)& Gieren et al. 1997    &      18.49 $\pm$ 0.05\\
\hline
\end{tabular}
\end{center}
\end{table}

\subsubsection{``The Pleiades problem''}

The Pleiades cluster  at 130 pc has been taken to be the first milestone of 
the distance work, since it has nearly solar abundance of heavy elements.
This cluster is already too far to obtain a reliable
parallax with the ground based 
observations, and its distance is estimated by tying it with nearby
stars with solar metallicity employing main sequence fitting of FGK 
dwarfs (e.g., van Leeuwen 1983).
The distance obtained this way agrees with an estimate via the
Hyades, the nearest cluster to which geometric distance is available
from the ground (Hanson 1980; van Altena et al. 1997), after a correction for 
large metallicity of the Hyades
(VandenBerg \& Bridges 1984).
It was then a natural exercise to confirm these estimates with a parallax
measured by the Hipparcos. The result showed that the Pleiades distance is
shorter by 0.25 mag (12\%) (van Leeuwen \& Hansen-Ruiz 1997, 
Mermilliod et al 1997)! 
This is summarised in TABLE 5.

Mermilliod et al.'s (1997) (see also de Zeeuw et al. 1997) have shown that
such a disagreement is seen not only for the Pleiades
but also for other open clusters 
to some degree.  
A noteworthy example is that the locus of the Praesepe ([Fe/H]=+0.095)
agrees with that of the Coma Ber ([Fe/H]=$-$0.065) {\it without} metallicity
corrections, while we 
anticipate the former to be 0.25 mag brighter due to higher metellicity.

This is a serious problem, since the
disagreement means that either our understanding of FGK dwarfs,
for which we have the best knowledge for stellar evolution, 
is incomplete, or the Hipparcos parallax contains systematic errors
(Pinsonneault et al. 1998; Narayanan \& Gould 1999). 
The origin is not understood yet.

\begin{table}[htb]
\begin{center}
\caption{Pleiades distance summary}
\begin{tabular}{lll}
\hline
Author/Method &  & Distance modulus \\
\hline
van Leeuwen (1983)&     & 5.57$\pm$0.08 \\
Ling\aa~(1987)     &     & 5.61   \\
\hline
Hyades (Perryman et al. 1998) & 3.33$\pm$0.01 &    \\
Pleiades$-$Hyades &             2.52$\pm$0.05  &   \\
metallicity correction &      $-$0.22$\pm$0.03  &    \\
             &                          &5.63$\pm$0.06 \\
\hline
van Leeuwen \& Hansen-Ruiz 1997& &5.32$\pm$0.05   \\
Mermilliod et al. 1997 &        &5.33$\pm$0.06   \\
van Leeuwen 1999  &             &5.37$\pm$0.07  \\
\hline
\end{tabular}
\end{center}
\end{table}

\subsubsection{Metallicity effects in the LMC Cepheid calibration}

The Cepheid distance to LMC is based on the calibration using open cluster
Cepheids, the distances to which are estimated by B star main sequence
fitting that ties to the Pleiades (Sandage \& Tammann 1968, Caldwell, 1983, 
Feast \& Walker 1987, Laney \& Stobie 1994).
Metallicity has been measured for some of these calibrator Cepheids
(Fry \& Carney 1997). The residual of the PL fit shows a strong metallicity
($Z$) dependence. This means {\it either} the 
Cepheid PL relation suffers from a large
$Z$ effect, or the distances to open clusters contain significant  
$Z$-dependent errors (Sekiguchi \& Fukugita 1998). 
A correction for this effect changes the distance to
LMC in either way, depending upon which interpretation is correct.

This metallicity dependence problem can be avoided if parallaxes are used
to find the distances to calibrator Cepheids. 
Attempts were made 
(Feast \& Catchpole 1997; Luri et al. 1998; Madore \& Freedman 1998)
using field Cepheids in the Hipparcos catalogue.
Unfortunately, Cepheid parallax data are so noisy (only 6 have errors 
less than 30\%) that they do not allow 
calibrations tighter than ladders.  
Another skepticism is that 2/3 of 
Cepheids in the nearby sample (e.g., 14/26 in the Feast-Catchpole sample)
are known to have companion stars, which would 
disturb the parallax (Szabados 1997).

\subsubsection{Red clump}

The OGLE group revived the use of the red clump (He burning stage of Population
I stars) as a distance indicator. Paczy\'nski \& Stanek (1998)
showed that the $I$ band luminosity of the red clumps depends
little on metallicity (see, Cole 1998, however), and 
gave a calibration using the Hipparcos parallax for nearby He 
burning stars. Udalski et al. (1998a) and 
Stanek et al. (1998) applied this to LMC, and 
obtained a distance modulus 18.1$\pm$0.1, much shorter than 
those from other methods.
This is a modern version of an analysis of
Mateo \& Hodge (1986), who reported 18.1$\pm$0.3.
We should also recall that earlier analyses using MS fitting of OB stars 
resulted in a short distance of 18.2$-$18.3 (Schommer et al. 1984;
Conti et al. 1986),
though somewhat dismissed 
in the modern literature.

\subsubsection{Detached eclipsing binaries}

Detached double-spectroscopic eclipsing binaries provide us with a unique
chance to obtain the distance in a semi-geometric way out to LMC or even
farther.
 From the information given by the light curve and velocity curve,
one can solve for the orbital parameters and stellar radii (Andersen 1991,
Paczy\'nski 1997; Bell et al 1993 for an earlier application to LMC
HV2226;
Torres et al. 1997 for an application to the Hyades). 
If surface brightness
of the two stars is known from colour or spectrum, one can obtain the 
distance as $d=(F/f)^{1/2}R_i$ where $F$ and $f$ are fluxes at the source and
the observer and $R_i$ is stellar radius. Guinan et al. (1998) 
applied this method to HV2274 in LMC
and derived $m-M=18.30\pm0.07$ with the aid of Kurucz' model atmosphere
to estimate surface brightness from the spectrum. 
Udalski et al. (1998b) claimed that the extinction used  
is too small by an amount of $\Delta E(B-V)=0.037$ mag based on OGLE
multicolour photometry. If we accept this correction the distance becomes
0.11 mag shorter, i.e., $m-M=18.19$. 

\subsubsection{RR Lyr problems}

In the first approximation the luminosity of RR Lyr is constant,
but in reality it
depends on metallicity. The dependence is usually expressed as
\begin{equation}
\langle M_V({\rm RR~Lyr})\rangle = a{\rm [Fe/H]} + b~.
\end{equation}
Much effort has been invested to determine $a$ and $b$.
The problem is again how to 
estimate the distance to RR Lyr. Unlike the case with Cepheids,
there are no unique ladders for the calibration, and a variety of 
methods have been used, of which the best known is the Baade-Wesselink method.
The calibration from the ground may be summarised as
\begin{equation}
\langle M_V({\rm RR~Lyr})\rangle= 0.2{\rm [Fe/H]} + 1.04. 
\end{equation}
With this calibration we are led to the LMC distance of
$m-M\simeq18.3$, as we saw in TABLE 1 above. 

The Hipparcos catalogue contains a number of field subdwarfs with
parallax. This makes a ladder available to calibrate RR Lyr
in globular clusters. 
Gratton et al. (1997) and Reid (1997) 
carried out this subdwarf fitting. Gratton et al. gave 
\begin{equation}
\langle M_V({\rm RR~Lyr})\rangle 
             = (0.22\pm0.09){\rm [Fe/H]}+0.76. 
\end{equation}
Their data are plotted in Figure 2, together with (4) and (5). 
Reid's result is also consistent.
This zero point, being brighter by 0.3 mag (at [Fe/H]=--1.8)
compared to (4), would bring the LMC distance to $m-M=18.5-18.6$.

There are a few analyses using the statistical parallax
for field RR Lyr in the Hipparcos catalogue. 
Fernley et al. (1998) reported that their 
halo RR Lyr lie almost exactly on the curve of (4), rather than (5),
and concluded a confirmation of the ground-based calibration.
This is also endorsed by an analysis of Gould \& Popowski (1998).

The distance to eponymous RR Lyr was measured by 
Hipparcos.
We see (Fig. 2) that RR Lyr itself does not fall on (5), but almost 
exactly on (4), although the error is fairly large. 
The uncertainties by 0.3 mag in the RR Lyr calibration 
translate to the LMC distance modulus 18.25$-$18.55. 

\subsubsection{Conclusions on the LMC distance}

The distance to LMC is uncertain as much as 0.4 mag (20\% in distance),
ranging from 18.20 to 18.60. The results are rather bimodal around
the two values close to the edges.
A geometric method with SN1987A ring echo initiated by
Panagia et al. 1991 
does not differentiate between these two values: the data are too noisy and
the result depends on the model of the light curve and 
emission lines that is adopted (Gould \& Uza 1997; Sonneborn et al. 1997;
see Fig. 8 of the latter literature for the data quality).
As we have seen in this section,  
recent observations with new techniques seem to
tip the list to the lower value. This is clearly a systematic effect,
so that we cannot simply take an `average of all observations'. Rather, 
we should leave both possibilities open.

\begin{figure}
\begin{center}
\centerline{{\epsfxsize=6cm\epsfbox{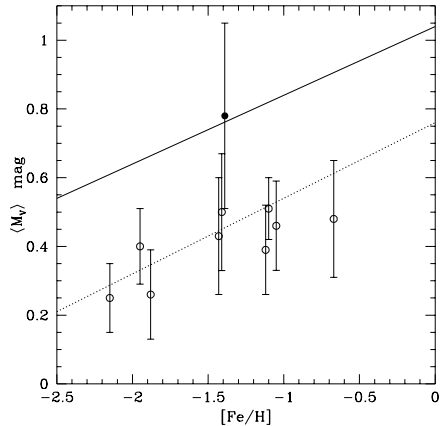}}}
\caption{Calibrations of RR Lyr. The open points are taken from Gratton et al.
(1997) with the dotted line indicating (5). The solid line is the
ground-based calibration (4). The solid point denotes the eponymous RR Lyr 
measured by the Hipparcos satellite.}
\end{center}
\vspace{-1cm}
\end{figure}

\subsubsection{Age of the globular clusters} 

The RR Lyr calibration is also crucial in the estimation of the age of 
globular clusters, since the stellar age is 
proportional to the inverse
of luminosity, i.e., inverse square of the distance. 
The modern evolution tracks of the main sequence agree reasonably well 
among authors. There are
some disagreements in colours around the turn-off point, 
largely depending on the treatment of
convection, but the luminosity is little affected 
(e.g., Renzini 1991; Vandenberg et al. 1996, especially their Fig. 1).
Absolute magnitude at the turn-off point $M^{TO}_V$ of the main sequence 
is hence a good indicator of the age, as 
(Renzini 1991),
\begin{equation}
\log t_9=-0.41+0.37M^{TO}_V-0.43Y-0.13{\rm [Fe/H]},
\end{equation}
in units of Gyr, or
\begin{equation}
\log t_9=-0.41+(0.37a-0.13){\rm [Fe/H]}+0.37[(M^{TO}_V-M_V^{\rm RR})+b]
]-0.43Y, 
\end{equation}
if (3) is included. The difference of the magnitudes between 
the turn-off point and RR Lyr $(M^{TO}_V-M_V^{\rm RR})$ varies little 
among 
clusters and is measured to be 3.5$\pm$0.1 mag (Buonanno et al. 1989;
see Chaboyer et al. 1996 for a compilation).
The metallicity dependence 
of the cluster age disappears if $a=0.35$, i.e., the globular
cluster formation is coeval (Sandage 1993b). 
Both (4) and (5), however, give $a\simeq0.2$,\footnote[2]{ 
A remark is made on a recent analysis of 
Kov\'acs \& Jurcsik (1996),
who obtained $a<0.19$ from a model-independent approach using the 
Fourier coefficients of the light curves that correlate with the 
metal abundance.}~  which 
indicates that metal-poor clusters appear older.

The dichotomous calibrations of RR Lyr obviously affect the age of 
globular clusters. Another large uncertainty is 
whether the age-metallicity correlation is
real, indicating metal-poor clusters formed earlier, 
or is merely
due to a systematic error, with the formation of globular cluster being 
coeval. The possibilities are four-fold:
\begin{eqnarray}
\begin{array}{cccc} 
b &(m-M)_{\rm LMC} &t_0({\rm noncoeval}) &t_0({\rm coeval}) \\ 
1.05 & 18.25 & 18{\rm Gyr} & 15{\rm Gyr}  \\ 
0.75 & 18.55 & 14{\rm Gyr} & 12{\rm Gyr}  \\
\nonumber
\end{array} 
\end{eqnarray}
In addition there are $\pm$10\% errors from various sources
(Renzini 1991; Bolte \& Hogan 1995; VandenBerg et al. 1996;
Chaboyer et al. 1996). Figure 3 shows the age of various clusters
from Gratton et al. (1997) and Chaboyer et al. (1998) both using
the calibration close to (5). The [Fe/H] dependence is apparent. 

The claims of Gratton et al. (1997), Reid (1997) and Chaboyer et al. (1998) 
for young universe
(11-12 Gyr) assume a coeval-formation interpretation together 
with the long RR Lyr calibration and take a mean of globular cluster ages.
Three other possibilities, however, are not excluded. 

\begin{figure}
\begin{center}
\centerline{{\epsfxsize=6cm\epsfbox{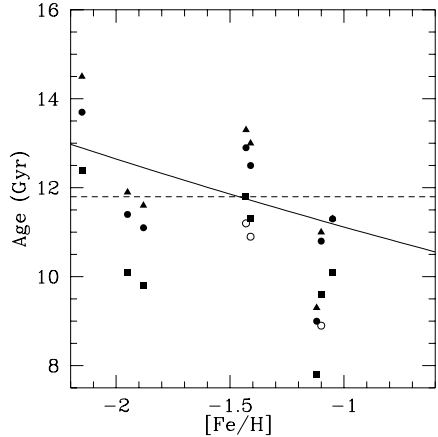}}}
\caption{Age of globular clusters as a function of [Fe/H]. 
The solid points are from Gratton et al. (1997) for three
different stellar evolution models. The open points are
from Chaboyer et al. (1998). The solid line shows (7) but offset by $-$0.06. 
The dashed line is 11.8 Gyr of Gratton et al.}
\end{center}
\vspace{-1cm}
\end{figure}

\subsection{Metallicity problems with Cepheids}

In most applications of the Cepheid PL relation, metallicity 
effects are neglected, motivated by
theoretical arguments that they will be very small. 
This results from double cancellations of the
metallicity dependences between core luminosity and atmosphere, as well as
between the effects of the helium abundance and of heavier elements. The
expected effect is (Stothers 1988, 
Iben\& Tuggle 1975; Chiosi et al. 1993)
\begin{equation}
\gamma_\lambda\equiv\partial M_\lambda/\partial{\rm [Fe/H]}\simeq \pm 0.05~{\rm dex~mag}^{-1}~~
\end{equation}  
for the $\lambda=V,I$ pass bands.
A new calculation of Sandage et al. (1999) gives $|\gamma_\lambda|<0.1$
for $ \lambda=B,V,I$.

When one is concerned with a 10\% systematic error in the cosmic distance
scale, 
the metallicity effect must be scrutinised.
If it were as large as
$-$0.5, say, the true Cepheid distance to normal 
spiral galaxies would be longer by 10\% relative to
LMC ([O/H]=$-$0.4). 
The calibrator SNe used in earlier papers (SN1937C, SN1972E, 1981B and
1990N) all reside in low $Z$ galaxies,
but recent additions include SNe in high $Z$ galaxies
(1989B, and notably 1998bu), thus the sample spans a wider 
metallicity range (see Figure 1 above).
There is now no relative difference in metallicity effects 
any more between the SBF and SNIa
calibrator samples (the offset is $\Delta{\rm [O/H]}<0.1$).   
Therefore, we cannot ascribe the difference in $H_0$ to the
metallicity effect of the Cepheid PL relation: the effect slightly reduces 
$H_0$ from both methods if the sign of $\gamma$ is negative.  

However, it is important to know the magnitude of $\gamma$.
Observationally, Freedman \& Madore (1990: FM) showed with the M31 data
that the metallicity dependence is small ($\gamma_{BVRI}=-0.32\pm0.21$). 
Gould (1994), however, reanalysed the same 
data and concluded it to be as large as  $\gamma=-0.88\pm0.16$. The EROS
collaboration derived $\gamma_{VI}=-0.44$ from a comparison between LMC
and SMC (Beaulieu et al. 1997). 
Kochanek (1997) suggested $\gamma_{VI}=-0.14\pm0.14$ from a global fit of
galaxies with Cepheid observations. 
The metallicity dependence for Galactic Cepheids discussed in 
section 2.3.2 corresponds to $\gamma_{VJHK}\approx-2$.
Kennicutt et al. (1998) pointed out that
the metallicity gradient of M31 used by  Freedman \& Madore is
a factor of three too large and argued that the above values should be
$\gamma_{BVRI}=-0.94\pm0.78$ (FM) and $-2.1\pm1.1$ (Gould).

Kennicutt et al. (1998) derived from HST observations of two fields in M101 
that $\gamma_{VI}=-0.24\pm
0.16$, which is the value currently adopted in metallicity dependence
analyses of the HST-KP group. If this is the true value, the effect on
the distance scale is of the order of 5$\pm$3\% ($H_0$ gets smaller).
I would emphasize, however, that independent confirmations
are necessary for this $\gamma$ value, since the M101 analysis 
is based only on $V$ and $I$ bands, and
the effect of extinction might not be completely disentangled.    
 

\subsection{Cross-check of the Cepheid distances}

\subsubsection{Tests with geometric methods}

NGC4258 (M106) is a Seyfert 2 galaxy with H$_2$O maser emission from 
clouds orbiting around a black hole of mass $4\times 10^7M_\odot$
located at the centre.
Precise VLBA measurements of Doppler velocities show that the motion
of the clouds is very close to Keplerian and is perturbed very little
(Miyoshi et al. 1995). 
A complete 
determination is made for the orbital parameters including
centripetal acceleration and a bulk proper motion of the emission system. 
This yields a geometric distance to NGC4258 to be $7.2\pm0.3$ Mpc
(Herrnstein et al. 1999). 

Maoz et al. (1999) measured the distance to NGC4258 using the conventional
Cepheid PL relation, and 
gave 8.1$\pm$0.4 Mpc with $(m-M)_{\rm LMC}=18.5$. This
distance is 13\% longer than that from the maser measurement. 
The short LMC distance would bring the Cepheid distance in a
perfect agreement with the geometric distance. 
This is, however, only one example,
and it can merely be a statistical effect: the deviation is only
twice the error, so it may happen
with a chance probability of 5\%.

\subsubsection{Further checks for M31}
 
A number of distance estimates are available for the nearest giant 
spiral M31, and they are shown in TABLE 6 (the underlined numbers
are the zero point).
Stanek \& Garnavich (1998) applied the red clump method to M31, 
and obtained $m-M=24.47\pm0.06$, which
agrees with the M31 modulus $24.44\pm0.10$ from the Cepheid based 
on the $(m-M)_{\rm LMC}=18.5$ calibration, whereas the same method 
gives 18.1 for the LMC distance. Namely, M31$-$LMC largely disagrees between
the two.  This discrepancy might be ascribed to a metallicity 
effect of either Cepheids or red clumps, or to the systematic
error of either indicator.  Mochejska et al. (1999) ascribed it
to the error of the Cepheid distance from a crowded stellar population.
On the other hand, tip of giant branch (TRGB) gives $(m-M)_{\rm M31-LMC}$ 
in agreement with the value derived from the Cepheid. 
The difference from PNLF is also consistent.
The value from RR Lyr, however, is consistent with that from red clumps
(the numbers in the Table is derived using (4) for the zero point).
The results are dichotomous again.

\begin{table}[htb]
\begin{center}
\caption{Relative distance of M31 to LMC}
\begin{tabular}{lllll}
\hline 
method & M31 & LMC & M31$-$LMC & refs.\\
\hline
Cepheid & 24.44$\pm$0.10 & \underline{18.5} & 5.94$\pm$0.10 & Ferrarese et al. 1999b \\
red clump & 24.47$\pm$0.06 & 18.07$\pm$0.04 & 6.40$\pm$0.07 & Stanek \& Garnavich 1998 \\
TRGB & 24.41$\pm$(0.19) & \underline{18.5} & 5.91$\pm$(0.19) & Ferrarese et al. 1999b \\
RR Lyr ($B$) & 24.50$\pm$(0.15) & 18.30 & 6.20$\pm$  & Pritchet \& vd Bergh 1989 \\
PNLF     & \underline{24.44} & 18.56$\pm$0.18 & 5.82$\pm$0.18 & Jacoby et al. 1990b \\
\hline
\end{tabular}
\end{center}
\end{table}

\subsection{Physical methods}

\subsubsection{Expansion photosphere model (EPM) for Type II supernovae}

This is a variant of the Baade-Wesselink method. If a supernova is a
black body emitter one can calculate source brightness from 
temperature; the distance can then be estimated by comparing 
source brightness with the observed flux. In SNeII atmosphere
the flux is diluted due to electron scattering opacity. If this 
greyness is calculated 
source brightness can be inferred. Schmidt, Kirshner \& Eastman (1992)
developed this approach and obtained the distances to SNeII in 
agreement with those from the ladder. The point is that
EPM gives absolute distance without zero point calibrations. 
The Hubble constant they obtained is 73$\pm$9 (Schmidt et al. 1994). 

A possible source of systematic errors is in the estimation of the temperature 
from the 
spectrum or colour. The SNeII physics also might not be uniform, 
as we see occasional large scatters in a 
cross-correlation analysis.

\subsubsection{Gravitational lensing time delay}

When quasar image is split into two or more by gravitational lensing, 
we expect the time delay among images, arising from different path 
lengths and gravitational potentials among image positions. The time delay between
images A and B takes the form
\begin{equation}
\Delta t={1+z_L\over H_0}\left({D_{OL}D_{OS}\over D_{LS}}H_0\right)
\left[{1\over2}(|\theta_A|^2-|\theta_B|^2-\Delta \phi|_{A-B})\right]
\end{equation}
where $\theta$ is the angular difference between the source and image,
$\Delta \phi$ is the difference in the potential 
and $D_{IJ}$ is the angular diameter distances. 
The time delay is observable if the source is variable,
and can be used to infer $H_0$ (Refsdal 1964). 
Crucial in this argument is
a proper modelling of the mass distribution of the deflector.
The $DD/D$ factor depends on $\Omega$ only weakly; its $\lambda$
dependence is even weaker.

The first case where $H_0$ is derived is with the 0957+561 lens system. 
The deflector
is complicated by the fact that a giant elliptical galaxy is embedded into a
cluster. Falco, Gorenstein \& Shapiro (1991) noted 
an ambiguity associated with a galaxy mass $-$ cluster mass separation,
which does not change any observed lens properties but affects the
derived Hubble constant. One way to resolve this degeneracy is to use the
velocity dispersion of the central galaxy (Falco et al. 1991;
Grogin \& Narayan 1996). Kundi\'c et al. (1997b), having resolved a 
long-standing uncertainty about the time delay, obtained 
$H_0=64\pm13$ employing the Grogin-Narayan model. 
Tonry \& Franx (1998) revised it to 
71$\pm$7 with their new velocity dispersion 
measurement near the central galaxy. 
More recently, Bernstein \& Fischer (1999) searched a wider
variety of models, also using weak lensing information to constrain the 
mass surface density of the cluster component, and concluded 
$H_0=77{+29\atop-24}$,
the large error representing uncertainties associated with the choice
of models.

The second example, PG1115+080, 
is again an unfortunate case. The deflector
is elliptical galaxy embedded in a Hickson-type compact group of 
galaxies (Kundi\'c et al. 1997a). 
Keeton \& Kochanek (1997) and Courbin et al. (1997) derived $(51-53)\pm15$
from the time delay measured by Schechter et al. (1997).
Impey et al. (1998) examined the dependence of the derived $H_0$ on the
assumption for the dark matter distribution, and found it to vary from 
44$\pm$4 (corresponding to $M/L$ linearly increasing with the distance)
to $65\pm5$ (when $M/L$ is constant over a large scale).
The latter situation may sound strange, but it seems not too unusual
for elliptical galaxies, a typical example being seen in NGC5128 
(Peng et al. 1998).

Recently, time delays have been measured for three more lenses, B0218+
357,
B1608+656 and PKS1830-211. B0218+357 is a rather clean, 
isolated spiral galaxy lens, and Biggs (1999) derived $H_0=69{+13\atop-19}$
(the central value will be 74 if $\Omega=0.3$)
with a simple galaxy model of a singular isothermal ellipsoid.
For B1608+656, Koopmans \& Fassnacht (1999) obtained 
64$\pm$7 for $\Omega=0.3$
(59$\pm$7 for EdS). For PKS1830-211, 
they gave 75${+18\atop-10}$
for EdS and 85${+20\atop-11}$ for $\Omega=0.3$
from the time delay measured by 
Lovell et al. (1998). 
More work is clearly needed to exhaust the class of models, but
these three lens systems
seem considerably simpler than the first two examples.
Koopmans \& Fassnacht concluded $74\pm8$ for low density cosmologies
($69\pm7$ for EdS) from four (excluding the second) lensing
systems using the simplest model of deflectors.
It is encouraging to find 
a good agreement with the values from the ladder argument, though the current
results from lenses are still less accurate than the ladder value.
It would be important to ask whether $H_0<60$ or
$>80$ is possible within
a reasonable class of deflector models.

\subsubsection{Zeldovich-Sunyaev effect}

The observation of the Zeldovich-Sunyaev (ZS) effect for clusters tells us
about the cluster depth (times electron density), which, when
combined with angular diameter (times electron density square)
from X ray observations, gives us the distance to the cluster provided
that cluster is spherical (Birkinshaw et al. 1991, Myers et al. 1997).  
This is often taken as a physical method 
to measure $H_0$. I give little weight to this method in these lectures, since
it is difficult to estimate the systematic errors.
The currently available results wildly vary from a cluster to a
cluster.
The most importnat is a bias towards elongation. 
None of the known clusters are quite spherical, and selection effects    
bias towards clusters elongated along the line of sight because of
higher surface brightness. This may happen even if one uses a large sample.
Additional systematics arise from the sensitivity of the ZS effect
to the cluster envelopes; one must resort to a model to correct for
this effect.
  
\subsubsection{Physical methods: summary}

Physical methods now yield the Hubble constant 
which can be compared with that from ladders. TABLE 7 presents 
a summary of $H_0$ from the physical methods. However, effort 
is still needed to determine systematic errors associated with the use 
of specific methods.
 
\begin{table}[htb]
\begin{center}
\caption{Hubble constant from ladders and physical methods}
\begin{tabular}{lll}
\hline 
method & $H_0$ & reference\\
\hline
{\bf ladders} & {\bf 71$\pm$7}($\times${\bf 0.95-1.15}) &  \\
physical: EPM &      73$\pm$9 & Schmidt et al. 1994  \\
physical: lensing (low $\Omega$) &  74$\pm$8 & Koopmans \& Fassnacht 1999  \\
 ~~~~~~~~~~~~~~~~~~~~~~(EdS) &    69$\pm$7  &  \\
physical: ZS  & (54$\pm$14) & Myers et al 1997  \\
\hline
\end{tabular}
\end{center}
\end{table}

\subsection{Conclusions on $H_0$}

The progress in determining the extragalactic 
distance scale has been dramatic. The ladders
yield values convergent within 10\%, which is compared to a factor 1.6
disagreement in the early nineties.  
A new uncertainty, however, becomes manifest in the Galactic
distance scale: there is a 15$-$20\% uncertainty in the distance to LMC.
Therefore, we may summarise
\begin{equation}
H_0=(71\pm 7)\times {1.15 \atop 0.95}
\end{equation}
as  a currently acceptable value of the Hubble constant. 
This agrees with that from HST-KP (Mould et al. 1999) up to the
uncertainty from the LMC distance, though we followed a different path
of argument. 
This allows
$H_0=90$ at the high end (if Tonry et al's SBF is weighted)
and 60 at the low end (if the SNeIa results are weighted). 
Note that $H_0$ from both EPM and gravitational lensing are
consistent with the ladder value for $(m-M)_{\rm LMC}=18.5$. With
the shorter LMC distance the overlap is marginal.


The short LMC distance will also cause trouble for the $H_0-$age
consistency. The LMC distance modulus of $m-M=18.25$ would raise the
lower limit of $H_0$ to 72, and increase the lower limit of age 
from $\approx$11.5 Gyr to $\approx$14.5 Gyr at the same time. 
There is then no solution for a $\lambda=0$
universe. With a non-zero $\lambda$, a unique 
solution is $H_0\simeq72$, $\Omega\simeq0.25$,  $\lambda\simeq0.75$
with coeval globular cluster formation (see Figure 6 below). 

In the future it is likely that more effort will be expended for geometric
methods. The great advantage is that it is free from errors 
arising from the chemical composition. In the surface brightness method, 
the chemical composition may still enter 
into the game, but its effect is tolerable and can even be reduced
to a negligible level by using near infrared observations. 

Ultimately, gravitational wave observations could provide us with a novel
method. For instance, for coalescing binary neutron stars the distance can be
calculated as $d\sim \nu^{-2}\varepsilon^{-1}\tau^{-1}$, where $\varepsilon$ is
metric perturbations, $\nu$ is the frequency and $\tau=\nu/\dot\nu$ 
is a characteristic
time of the collapse (Schutz 1986). The
position of the object may be difficult to infer, 
but there might be a gamma ray burst associated with the
coalescence.

\section{The density parameter}

\subsection{Model-independent determinations} 

\subsubsection{Luminosity density $\times$ $\langle M/L\rangle$}

The mass density can be obtained by multiplying the luminosity density
with galaxy's average mass to light ratio $\langle M/L\rangle$. 
The local luminosity density, evaluated by integrating the luminosity function,
is reasonably well converged to
${\cal L}_B=(2.0\pm 0.4)\times 10^8 h L_\odot$ Mpc$^{-3}$ from
many observations.
The $M/L_B$ of galaxies generally increases with the scale. 
When the mass is integrated to
$\approx 100$ kpc, a typical $M/L_B$ is about $(100-200)h$ 
in solar units, 
and it may still
increase outward (e.g., Faber \& Gallagher 1979; Little \& Tremaine 1987; 
Kochanek 1996; Bahcall et al. 1995; Zaritsky et al. 1997). 
The virial radius in a spherical collapse model is
$r=0.13~{\rm Mpc}~\Omega^{-0.15}[M/10^{12}M_\odot]^{1/2}_{<100{\rm kpc}}$.
If the dark matter distribution is isothermal within the virial radius,
the value of $M/L_B$ inside the virial radius is $(150-400)h$ for $L^*$
galaxies.  
This is about the value of $M/L_B$ for groups and clusters, 
$(150-500)h$. 
Multiplying the two values we get
$
\Omega=0.20\times 2^{\pm1}. 
$
See also Fukugita, Hogan \& Peebles (1998) for variants of this argument.

Carlberg et al. (1996, 1997a) tried to make the argument more quantitative
using their cluster sample and a built-in field galaxy sample. They 
estimated $M/L_r\simeq (210\pm60)h$ for field galaxies from the 
cluster value $(289\pm50)h$. Their   
luminosity density of field galaxies is   
 ${\cal L}_r=(1.7\pm 0.2)\times 10^8 h L_\odot$ Mpc$^{-3}$,
and therefore $\Omega_0=0.19\pm0.06$. Note that  
$M/L_B\simeq1.4\times M/L_r$ in solar units for the respective pass bands.

The important assumption for these calculations is the absence
of copious matter outside the clusters. This is a question difficult
to answer, but the observation of weak lensing
around the clusters indicate that the distributions of dark mass
and galaxies are similar at least in the vicinity of clusters
(Tyson \& Fischer 1995; Squires et al. 1996).

Some attempts have also been made to estimate the mass on a supercluster scale.
Small et al. (1998) inferred $M/L_B\simeq 560h$ for the Corona Borearis 
supercluster, by applying
the virial theorem (inspired by an $N$ body simulation). On the other
hand, Kaiser et al. (1998) estimated $M/L_B\simeq 250$ from a mesurement
of the gravitational shear of weak lensing caused by a supercluster MS0302+17
\footnote[3]{They suggest $\Omega\simeq 0.04$ on the basis that only
early-type galaxy population traces the mass distribution and the luminosity
density is multiplied by the fraction of early-type galaxies (20\%). 
It seems possible that late type galaxies reside in low density 
regions, causing only a small shear, which is buried in noise, 
and escaped from the measurement.};
the result is not well convergent, but it seems unlikely that $\Omega$ 
is larger than 0.5.

\subsubsection{$H_0$ versus cosmic age}

For $H_0\ge 60$, the age is 10.9 Gyr for the EdS universe. This is too
short. $\Omega$ must be smaller than unity. If we take $t_0>11.5$ Gyr 
$\Omega<0.7$. The limit is weak, but the significance is 
that EdS universe is nearly excluded.

\subsubsection{Type Ia supernova Hubble diagram}

The type Ia supernova Hubble diagram now reaches $z\simeq 0.4-0.8$.
It can be used to infer the mass density parameter and the cosmological
constant. As we discuss later (section 4.1)
the observation favours a low $\Omega$
and a positive $\lambda$.
If we take their formal errors, $\Omega<0.1$ is allowed only at three 
sigma for a zero $\lambda$ universe 
(Riess et al. 1998; Perlmutter et al. 1999).
A zero $\lambda$ open universe may not be excluded yet if some allowance
is taken for systematic effects, but EdS geometry
is far away from the observation.
The best favoured value is approximately, 
\begin{equation}
\Omega\approx 0.8\lambda-0.4~.
\end{equation}

\subsubsection{Baryon fractions}

A cluster is a virialised object with the cooling time scale longer 
than the dynamical time scale, and hence the physics is governed only by 
gravity (except for cooling flows in high density regions).
The gas in clusters is shock heated
to the virial temperature  
$T\simeq 7\times10^{7}(\sigma/1000 {\rm km~s}^{-1})^2$ K, and thus emits
X rays by thermal bremsstrahlung. 
 From the luminosity and temperature of X rays one can 
infer the mass of the X ray emitting gas. It has been known that the  
gas amounts to a substantial fraction of the dynamical mass, which
means that baryons reside more in the gas than in stars by an order 
of magnitude
(Forman \& Jones 1982). The argument was then elaborated by
White et al. (1993b) 
based on ROSAT observations.
 From 19 clusters White \& Fabian
(1995) obtained $M_{\rm gas}/M_{\rm grav}=0.056h^{-2/3}$, where $M_{\rm grav}$
is the dynamical mass. By requiring that the cluster
baryon fraction agrees with $\Omega_B/\Omega$
in the field,
we have 
$\Omega=0.066h^{-1/2}\eta_{10}=0.39(\eta_{10}/5)$,
where $\eta_{10}$ is the baryon to photon ratio in units
of $10^{-10}$ and the last number assumes $h=0.7$.

An independent estimate 
is made from the Zeldovich-Sunyaev effect observed
in clusters (Myers et al. 1997; Grego et al. 1999):  
$M_{\rm gas}/M_{\rm grav}=0.082h^{-1}$
is translated to $\Omega=0.044h^{-1}\eta_{10}=0.31(\eta_{10}/5)$.  

If we insert a probable value of the baryon to photon ratio
from primordial nucleosynthesis calculations, $\eta_{10}=3-5$, we have
$
\Omega=0.2-0.4. 
$

\subsubsection{Peculiar velocity - density relation}

This is one of the most traditional methods to estimate the cosmic mass
density. The principles are spelled out by Peebles (1980).
There are two basic tools depending on the scale. For small scales
($r<1$ Mpc) the perturbations developed into a non-linear regime, and 
the statistical equilibrium argument is invoked for ensemble averages 
that the peculiar acceleration 
induced by a pair of galaxies is balanced by relative motions 
(cosmic virial theorem).   
For a large scale ($r>10$ Mpc), where perturbations are still
in a linear regime,
the basic equation is 
\begin{equation}
\nabla\cdot \vec{v}+H_0\Omega^{0.6}\delta=0
\end{equation}
%
\noindent
with $\delta$ the density contrast. The contribution from a cosmological
constant is negligible.
The problem inherent in all arguments involving velocity is 
the uncertainty regarding the extent to which 
galaxies trace the mass distribution (biasing),
or how much mass is present far away from galaxies. 

\paragraph{Small-scale velocity fields:}

The status is summarised in Peebles (1999), where he has concluded 
$\Omega(10{\rm kpc}\lsim r \lsim 1{\rm Mpc})=0.15\pm0.10$ from
the pair wise velocity dispersion (with samples excluding clusters) 
and the three point correlation function of
galaxies via a statistical stability argument.
Bartlett \& Blanchard (1996) argued
that it is possible to reconcile the observed velocity dispersion
with $\Omega\sim1$
if one assumes galactic halo extended beyond $>300$kpc.  As Peebles (1999)
argued, however, the halo is unlikely to be extended that much 
as indicated by the agreement of MW's mass at 100-200kpc and the 
mass estimate for MW+M31 in the Local Group.

Beyond a 10 Mpc scale, linear perturbation theory applies. An integral form
of (12) for a spherical symmetric case 
($v/H_0r=\Omega^{0.6}\langle\delta\rangle/3$) applied to the Virgocentric 
flow
gives $\Omega\simeq 0.2$ for $v\simeq 200-400 {\rm km~s}^{-1}$ and 
$\langle\delta\rangle\sim 2$, assuming no biasing 
(Davis and Peebles 1983). 
Recently, Tonry et al. (1999) argued that 
the peculiar velocity ascribed to Virgo cluster is only 140 km s$^{-1}$,
while the rest of the peculiar velocity flow is attributed to the
Hyd-Cen supercluster and the quadrupole field. 
For this case $\Omega\simeq0.06$.  We may have $\Omega\sim1$ only when 
half the mass is well outside the galaxies. 

Peebles (1995) argued that the configuration and kinematics of 
galaxies are grown following the least action principle
from the nearly homogeneous primeval mass distribution.
Applying this formalism to Local Group galaxies,
he inferred $\Omega=0.15\pm0.15$. On the other hand, 
Branchini \& Carlberg (1994) and 
Dunn \& Laflamme (1995) argued that this conclusion is not tenable 
if mass is distributed smoothly outside galaxies as in $\Omega=1$ CDM models. 
This seems, however, not very likely 
unless
mass distribution is extended over 10 Mpc scale (Peebles 1999).

\paragraph{Large scale velocity fields:}

There are a few methods to analyse the large-scale velocity fields
based on (12). The direct use of (12) is a comparison of the density
field derived from redshift surveys with measured peculiar velocities.
Alternatively, one may use the density field reconstructed from 
observed velocity
field for comparison with the actual density field, 
as in the {\stt POTENT} programme
(Dekel et al. 1990). A variant of
the first method is to observe the anisotropy in redshift space
(redshift distortion) (Kaiser 1987). As linear theory applies, $\Omega$
always appears in the combination $\beta= \Omega^{0.6}/b$ where $b$
is a linear biasing factor of galaxies against the mass distribution
and can be inferred through non-linear effects. 
Much effort has been invested in such analyses (see e.g.,
Strauss \& Willick 1995; Dekel et al. 1997; Hamilton 1998),
but the results are still controversial.
The value of $\Omega^{0.6}/b$ derived from many analyses varies from
0.3 to 1.1, though we see a general trend to favour a high value.
Notably, the most recent  {\stt POTENT} analysis using
the Mark III compilation of velocities (Willick et al. 1997) indicates 
a high density universe $\Omega=0.5-0.7$, and $\Omega>0.3$ only at 
a 99\% confidence level (Dekel et al. 1999).

The difficulty is that one needs
accurate information for velocity fields, for which an accurate estimate
of the distances is crucial. Random errors of the distance indicators
introduce large noise in the velocity field. This seems particularly serious
in the {\stt POTENT} algorithm, in which the 
derivative $\nabla\cdot \vec{v}/\Omega^{0.6}$
and its square are numerically computed; this 
procedure enhances noise, especially for a small $\Omega$. 
The difficulty of inferring large scale velocity
field may also be represented by the `great attractor problem'.
Lynden-Bell et al. (1988) found a large-scale velocity field towards the 
Hyd-Cen supercluster, but also argued that this  supercluster is also
moving towards the same direction attracted by a `great (giant) attractor'. 
With Tonry et al.'s (1999) new estimate of the distance using SBF, 
this velocity field is modest, and Hyd-Cen itself serves as 
the great attractor that pulls the Virgo cluster,
with the conclusion that $\Omega$ is small.

\subsection{Model-dependent determinations}

The following derivations of the mass density parameter are based on
the hierarchical clustering model 
of cosmic structure formation assuming the cold dark matter dominance.
The extraction of $\Omega$ is, therefore, indirect, but on the
other hand, it is reasonable to appeal to such models since $\Omega$ is
the parameter that predominantly controls structure formation. 
Note that CDM model is the only model known today that successfully predicts 
widely different observations, yet there are no observations strong
enough to refute its validity. We do not discuss 
results from cosmological models where physical processes
other than gravity play a major role.

\subsubsection{Shape parameter of the transfer function}

The initial perturbations of the density fluctuation 
$P(k)=|\delta_k|^2\sim k^n$ 
receive a modification as $P(k)=|\delta_k|^2\sim k^n T(k)$ as they grow, 
where $T(k)$ is called the transfer function. 
Fluctuations of a small scale that enter the horizon in the radiation dominant
epoch do not grow for a while, till the universe becomes matter dominated.
The transfer function $T(k)$ thus damps for small scales as  $\sim k^{-4}$,
whereas it stays close to unity for long-wave lengths. 
The transition region is controlled by a parameter
$k\sim 2\pi/ct_{\rm eq}$, $ct_{eq}$ being the horizon size at the
time of matter-radiation equality, i.e., a characteristic length of 
$6.5(\Omega h)^{-1}h^{-1}$ Mpc. The parameter 
$\Gamma=\Omega h$ determines the behaviour of the transfer function and 
is called the shape parameter. 
To give a sufficient power to several 
tens of Mpc, $\Gamma$ must be as small as 0.2
(Efstathiou et al. 1990). This small value ($\Gamma=0.15-0.25$) is
supported by later analyses (e.g., Peacock \& Dodds 1994; Eke et al.
1998).

\subsubsection{Evolution of the rich cluster abundance}

The cluster abundance at $z\approx 0$ requires  
the rms mass fluctuation 
$\sigma_8=\langle(\delta M/M)\rangle^{1/2}|_{r=8h^{-1}{\rm Mpc}}$ to
satisfy (White et al. 1993a; Eke et al. 1996; Pen 1998; Viana \& Liddle 1999;
see also Henry \& Arnaud 1991) 

\begin{equation}
\sigma_8\approx 0.6\Omega^{-0.5}~.
\end{equation}

The evolution of the cluster abundence is sensitive to $\sigma_8$ in early
epochs of growth for a given mass; it is $z\gsim0.3$ for rich clusters.
The rich cluster abundance at $z\sim0.3-1$,  when compared with 
that at a low $z$,
determines both $\sigma_8$ and $\Omega$ (Oukbir \& Blanchard 1992). 
Carlberg et al. (1997b) derived $\Omega=0.4\pm0.2$, and 
Bahcall \& Fan (1998) obtained 
$\Omega=0.2{+0.3 \atop -0.1}$ corresponding to a slow growth of the abundance. 
On the other hand, Blanchard \& Bartlett
(1998) obtained $\Omega\simeq 1$ from a more rapid growth. 
A high value is also claimed by
Reichart et al. (1999), while Eke et al. (1998)
reported $\Omega=0.43\pm0.25$ for an open, 
and $\Omega=0.36\pm0.25$ for a flat universe.

The controversy among authors 
arises from different estimates of the cluster mass at
high $z$. This is a subtle effect, since the mass varies little
over the range of relevant redshift, while the cluster number density
evolution is sufficiently rapid at fixed mass (Pen 1998). 
At low $z$ we have an established mass temperature relation,
and the cluster mass is securely 
estimated (Henry \& Arnaud 1991). At high $z$, however, such direct 
information is not available.  Blanchard \& Bartlett and Eke et al. 
used mass temperature relations as a function of $z$ 
derived from hydrodynamic simulations. Reichart et al. used an 
extrapolated mass X-ray luminosity relation. Bahcall and Fan 
used more direct estimates of the cluster mass at higher $z$
for three clusters. A change of a factor of two in the mass estimate
would modify the conclusion. 

\subsubsection{Cluster abundance versus the COBE normalisation} 

There are a number of ways to infer $\sigma_8$ from galaxy clustering
and peculiar velocity fields. The problem with the information from
galaxy clustering is that it involves an unknown biasing factor, 
which hinders us from determining an accurate $\sigma_8$. 
The velocity data are susceptible to noise from the distance 
indicators. Therefore, the cluster abundance discussed above 
seems to give us a unique method to  
derive an accurate estimate of $\sigma_8$ for a low $z$ universe. 
Another place  
we can extract an accurate  $\sigma_8$ is the fluctuation power
imprinted on  cosmic microwave background radiation (CBR) 
anisotropies. Currently
only the COBE observation (Bennett et al. 1996) gives sufficiently accurate 
$\sigma_8=\sigma_8(H_0, \Omega, \lambda, \Omega_B, ...)$. Assuming the
model transformation function, the matching
of COBE $\sigma_8$ with that from the cluster abundance gives a 
significant constraint
on cosmological parameters $\Omega=\Omega(H_0, \lambda)$
(Efstathiou et al. 1992; Eke et al. 1996)
Figure 4 shows allowed regions for two
cases, open and flat universes, assuming a flat perturbation spectrum $n=1$
and ignoring possible tensor perturbations.

The transfer function is modified if $n\ne1$. The 
possible presence of the tensor perturbations in CBR anisotropies causes
another uncertainty.
The COBE data alone say $n$ being between 
0.9 and 1.5 (Bennett et al. 1996), but the allowed range is
narrowed to $n=0.9-1.2$ if supplemented by  
smaller angular-scale CBR anisotropy data 
(Hancock et al. 1998; Lineweaver 1998; Efstathiou et al. 1998;
Tegmark 1999). The presence of the tensor
mode would make the range of $n$ more uncertain 
as well as it reduces the value of 
$\sigma_8$. The limit of $n$ when the tensor mode is maximally allowed
is about $<1.3$\footnote[4]{In Tegmark's analysis $n<1.5$ is quoted as 
an upper bound, but this is obtained by making $\Omega_B$ (and $H_0$)
a free parameter.
If one would fix the baryon abundance,
the allowed range is narrower, $n\lsim1.3$.}. Notwithstanding 
these uncertainties,
$\Omega>0.5$ is difficult to reconcile with the matching condition.
On the other hand, a too small $\Omega$ ($\lsim0.15$) is not consistent
with the cluster abundance.


\begin{figure}
\begin{center}
\centerline{{\epsfxsize=6cm\epsfbox{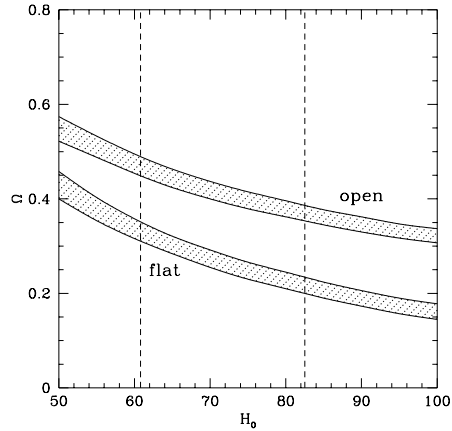}}}
\caption{Parameter regions allowed by matching the rms fluctuations
from COBE with those from the cluster abundance. A flat spectrum
($n=1$) is assumed and the tensor perturbations are neglected.
The lower band is for a flat universe, and the upper one for a universe
with $\Lambda=0$.}
\end{center}
\vspace{-1cm}
\end{figure}

\subsubsection{Power spectrum in nonlinear galaxy clustering}

Peacock (1997) argued that the power spectrum in a small scale region
($k^{-1}<3 h^{-1}$ Mpc),
where nonlinear effect is dominant, shows more power
than is expected in $\Omega=1$ cosmological models. He showed that
the excess power is understood if the mass density is   
$\Omega\approx0.3$.
   
\subsubsection{CBR anisotropy harmonics}

The $\ell$ distribution of the CBR harmonics $C_\ell$ depends on
many cosmological parameters. Precise measurements of
the harmonics will allow 
an accurate determination of the cosmological 
parameters up to geometrical degeneracy  
(Zaldarriaga et al. 1997;  Efstathiou \& Bond 1999; Eisenstein et al. 1999). 
At present the data do not give any constraint on
$\Omega$, but on some combination of $\Omega$ and $\lambda$;
so we defer the discussion to the next section.



\section{Cosmological constant}

Currently three tests yield useful results on the problem as to
the existence of the cosmological constant: 
(i) the Hubble diagram
for distant type Ia supernovae; (ii) gravitational lensing frequencies 
for quasars; (iii) position of the acoustic peak in the
harmonics of CBR anisotropies.

\subsection{Type Ia supernova Hubble diagram}

The luminosity distance receives a cosmology dependent correction as
$z$ increases; in a way $\Omega$ pulls down $d_L$ and $\lambda$ 
pushes it up. (In the first order of $z$ the correction enters
in the combination of $q_0=\Omega/2-\lambda$,  so this is often 
referred to as a $q_0$ test.) The discovery of two groups  
that distant supernovae are fainter than are expected from the
local sample, even fainter than are expected for $q_0=0$, 
points to the presence of $\lambda>0$ (Riess et al. 1998; Schmidt et al. 1998;
Perlmutter et al. 1999).

The general difficulty with such a Hubble diagram analysis is that one has to 
differentiate among a few interesting
cosmologies with small differences of brightness. 
For instance, at $z=0.4$ where many supernovae are observed,
the difference is $\Delta m=0.12$ mag between $(\Omega,\lambda)=
(0.3,0.7)$ and $(0, 0)$, and   
$\Delta m=0.22$ from  $(0, 0)$ to $(1.0,0)$. Therefore, an accuracy
of ($\lsim$5\%) must be attained including systematics to conclude the
presence of $\Lambda$.
On the other hand, there are a number of potential sources of errors: 

\noindent
(i) K corrections evaluated by integrating spectrophotometric data
    that are dominated by many strong features;

\noindent
(ii) relative fluxes at the zero point (zero mag) across the colour bands;

\noindent
(iii) dust obscuration in a host galaxy;

\noindent
(iv) subtraction of light from host galaxies;

\noindent
(v) identification of the maximum brightness epoch, and estimates of the
     maximum brightness including a $\Delta m_{15}$ correction;

\noindent
(vi) selection effects (for high $z$ SNe);

\noindent
(vii) evolution effects.

\begin{table}[htb]
\begin{center}
\caption{Estimates of maximum brightness on SNe: 1997 vs. 1999
from Perlmutter et al. (1997; 1999).}
\begin{tabular}{llll}
\hline 
SN & 1997 value & 1999 value& difference\\
\hline
SN1992bi & (23.26$\pm$0.24) & 23.11$\pm$0.46 & (0.15) \\
SN1994H  &  22.08$\pm$0.11  & 21.72$\pm$0.22 & 0.36   \\
SN1994al & 22.79$\pm$0.27   & 22.55$\pm$0.25 & 0.24   \\
SN1994F  & (21.80$\pm$0.69)  & 22.26$\pm$0.33 & \hskip-1mm ($-$0.58) \\
SN1994am &  22.02$\pm$0.14  & 22.26$\pm$0.20 & \hskip-1mm $-$0.24 \\
SN1994G & 22.36$\pm$0.35 & 22.13$\pm$0.49 & 0.23 \\
SN1994an & 22.01$\pm$0.33 & 22.58$\pm$0.37 & $-$0.57\\
\hline
\end{tabular}
\end{center}
Note: The numbers in the parentheses are not used in the final result of the
1997 paper.
\end{table}

Except for (vii), for which we cannot guess 
much\footnote[5]{Riess et al. (1999)
showed that the rise time is different between low $z$ and high $z$
samples, indicating some evolution of the SNIa population. The effect
on the cosmological parameter is not clear.}, the most important seems to be
combined effects of (i), (ii) and (iii). It is not easy a task to reproduce
a broad band flux by integrating over spectrophotometric data convoluted
with filter response functions, especially when spectrum contains strong
features. (Even for the spectrophotometric standard stars, the synthetic
magnitude contains an error of 0.02$-$0.05 mag,
especially when the colour band involves the Balmer or Paschen 
regions.) Whereas Perlmutter et al. assigns 0.02 mag to 
the error of (i) [and (ii)], a comparison of the two values of 
estimated maximum brightness
in their 1997 paper (Perlmutter et al. 1997, 
where they claimed evidence for a high $\Omega$ universe) 
and the 1999 paper (TABLE 8) 
shows a general difficulty in the evaluation of 
the $K$ correction (the difference dominantly comes from different
$K$ corrections). Schmidt et al. claim that their K correction errors
are 0.03\% mag. Dust obscuration (iii) is also an important source of
errors, since the error of (i)+(ii) propagates to $E(B-V)$ and 
is then amplified with the $R$ factor. So a 0.02 mag error in 
colour results in a 0.06 mag error in $A_V$. 

We note that each SN datum contains $\pm$0.2 mag (20\%)
error. The issue is whether this error is almost purely of random nature
and systemtics are controlled to a level of $\lsim$0.05.

\subsection{Gravitational lensing frequencies for quasars}

The gravitational lensing optical depth is given by

\begin{equation}
d\tau=\mu FH_0^3 (1+z_L)^3
\left[{D_{OL}D_{LS}\over D_{OS}}\right]^2{dt\over dz}dz 
\end{equation}

\noindent
where $F=\langle 16\pi^3n_g\sigma^4_gH_0^{-3}\rangle$, and $\mu$ is a 
magnification factor. The cosmological factor
in (14) is very sensitive to the cosmological constant, when it 
dominates (Fukugita \& Turner 1991). 
$F$ is the astrophysical factor that depends on the galaxy number
density $n_g$ and the mass distribution of galaxies, which is usually
assumed to be a singular isothermal sphere with velocity dispersion
$\sigma_g$. Figure 5 shows a typical
calculation for the expected number of strong lenses for 504 quasars
of the HST Snapshot Survey (Maoz et al. 1993) sample: 
the observed number is
5 (4 if 0957+561 is excluded). 
The curve shows a high sensitivity to $\lambda$ for $\lambda>0.7$,
but in contrast a nearly flat dependence for a lower $\lambda$. It is 
likely that $\lambda>0.8$ is excluded. On the other hand, a more
stringent limit is liable to be elusive. 
Fifty percent uncertainty in the $F$ factor, 
say, would change largely a limit on, or a likely value of, $\lambda$.

\begin{figure}
\begin{center}
\centerline{{\epsfxsize=6cm\epsfbox{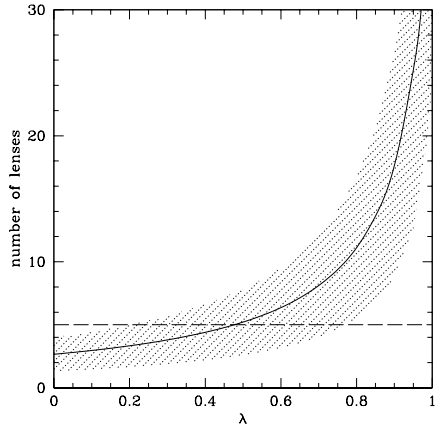}}}
\caption{Gravitational lensing frequencies as a function of $\Lambda$
in a flat universe. The expecetd number is given for 504 quasars
of the HST Snapshot Survey sample. The shade means the region within
a $\pm$50\% uncertainty. The observed number is 5 (dashed line).}
\end{center}
\vspace{-1cm}
\end{figure}

In order to acquire information for a smaller $\lambda$, an
accurate estimate is essential for the $F$ factor, which receives the following
uncertainties in: (1) the luminosity density and the fraction of early-type
galaxies (the lensing power of E and S0 galaxies is much higher than that
of spirals, and $F$ is roughly proportional to the luminosity density of
early-type galaxies); (2) $\sigma_g$-luminosity relation  
(Faber-Jackson relation); (3) the relation
between $\sigma$(dark matter) and $\sigma$(star); (4) the model
profile of dark haloes, specifically the validity of the
singular isothermal sphere approximation (note that dark matter distributions
seem more complicated in elliptical galaxies than in spiral galaxies,
see Fukugita \& Peebles 1999); (5) the core radius
which leads to a substantial reduction in $d\tau$;
(6) selection effects of the observations; (7) dust obscuration; (8)
evolution of early-type galaxies. 

There are continuous efforts 
for nearly a decade that have brought substantial improvement in 
reducing these uncertainties (Maoz \& Rix 1993; Kochanek 1996; 
Falco et al. 1998). Nevertheless,
the issue (1) still remains as a cause of a large uncertainty.
While the total luminosity density is known to an uncertainty of 20\% or so,
the fraction of early type galaxies is more uncertain. 
It varies from 0.20 to 0.41 depending on the literature. 
Including other items, it is likely that an estimate of $F$ has a
50\% uncertainty. For the curve in Figure 5 a change of $F$ by $\pm$50\% 
brings the most likely value of $\lambda$ to 0.75 or 0.2. 

Kochanek and collaborators have made detailed
considerations on the above uncertainties, and
carried out elaborate statistical analyses. 
In their latest publication they concluded 
$\lambda<0.62$ at 95\% confidence level from an optical sample
(Kochanek 1996). 
They took the fraction
of early-type galaxies to be 0.44 and assigned a rather small 1$\sigma$
error. (The predicted frequency comes close to the upper envelope of
Fig. 5, and the observed number of lenses in the HST sample is taken to be 4).
If one would adopt a smaller early-type fraction, the limit is
immediately loosened by a substantial amount. 
Since the uncertainty is dominated by systematics rather than
statistical, it seems dangerous to give significance to statistics. 
Statistical significance depends on artificial
elements as to what are assumed in the input.
A similar comment also applies to the recent work
claiming for a positive $\lambda$ (Chiba \& Yoshii 1997;
Cheng \& Krauss 1998). 
I would conclude a conservative limit being $\lambda<0.8$.

\subsection{Harmonics of CBR anisotropies}

This is a topic discussed repeatedly in this Summer Institute, 
so I will only briefly mention it for completeness. 
The positions of the acoustic
peaks are particularly sensitive to $\Omega$ and $\lambda$, and
even low accuracy data available at present lead to a meaningful
constraint on a combination of  $\Omega$ and $\lambda$.

The first acoustic peak appears at $\ell=\pi$(the distance
to the last-scattering surface)/(the sound horizon)
(Hu \& Sugiyama 1995). Its position
$\ell_1$ 
is approximated as
\begin{equation}
\ell_1\simeq 220\left({1-\lambda\over\Omega}\right)^{1/2}~,
\end{equation}
for the parameter range that concerns us. 
This means that the position of the acoustic peak is about 
$\ell\simeq 220$ if $\Omega+\lambda=1$, but it 
shifts to a high $\ell$ as $\Omega^{-1/2}$ if $\lambda=0$. 
On the other hand, there is little power to determine $\Omega$
separately from $\lambda$, unless full information of $C_\ell$ 
is used.
The harmonics $C_\ell$ measured at small angles revealed the
acoustic peak (Scott et al. 1996), and its position 
favours a universe not far from flat (Hancock et al. 1998).
More exhaustive analyses of Lineweaver (1998) and 
Efstathiou et al. (1999) show a limit
$\Omega+\lambda/2>0.52$ (1$\sigma$). (The contours of the confidence level
fall approximately on the curve given by (15) with $\ell_1$=constant.) 
This means that a zero
$\Lambda$ universe is already marginal, 
when combined with $\Omega$ from other arguments.
If a flat universe is chosen from CBR, a non-zero $\Lambda$ will be
compelling.

\begin{table}[htb]
\begin{center}
\caption{Summary of $\Omega$ and $\lambda$. }
\begin{tabular}{llll}
\hline 
method & $\Omega_0$ & $\Lambda$? & model used?\\
\hline
$H_0$ vs $t_0$ &  $<0.7$ \\
luminosity density +M/L & 0.1-0.4 & \\
cluster baryon fraction & 0.15-0.35& \\
SNeIa Hubble diagram & $\leq 0.3$ & $\lambda\approx 0.7$ \\
small-scale velocity field (summary) & $0.2\pm0.15$ & \\
 ~~~~~~~~~~~~~~ (pairwise velocity)  & $0.15\pm0.1$ & \\
 ~~~~~~~~~~~~~~ (Local Group kinematics) & $0.15\pm0.15$ & \\
 ~~~~~~~~~~~~~~ (Virgocentric flow) & $0.2\pm0.2$ \\
large-scale vel field & 0.2$-$1 & \\
cluster evolution (low $\Omega$ sol'n) & 0.2${+0.3\atop -0.1}$  & & ~~~yes\\
 ~~~~~~~~~~~~~~~~~~~~~~ (high  $\Omega$ sol'n) & $\sim$1  & & ~~~yes\\
COBE-cluster matching & 0.35-0.45 (if $\lambda=0$) &  & ~~~yes \\
                      & 0.20-0.40 (if $\lambda\ne0$) &  & ~~~yes \\
shape parameter $\Gamma$ & $0.2-0.4$ &  & ~~~yes\\
CBR acoustic peak \hfil& free (if flat) & $\gsim1-2\Omega$ & ~~~yes \\                       
                  & $>0.5$  (if open) &   & ~~~yes \\
gravitational lensing & & $\lambda<0.8$ \\
\hline
summary   & 0.15$-$0.45  (if open) &   &  \\
          & 0.2$-$0.4~~  (if flat) &   &  \\
          &                        & 0.6$-$0.7(?) & \\    
\hline
\end{tabular}
\end{center}
\end{table}

\section{Conclusions}

The status of $\Omega$ and $\lambda$ is summarised in TABLE 9.
We have a reasonable convergence of the $\Omega$ parameter towards
a low value $\Omega=0.15-0.4$.  The
convergence of $\Omega$ is significantly better with the presence
of the cosmological constant that makes the universe flat. 
Particularly encouraging is that the $\Omega$ parameters
derived with the aid of structure formation models agree with each other.
This is taken to be an important test for the
cosmological model, just as in particle physics when
many different phenomena are reduced to a few convergent
parameters to test the model. 
There are yet a still highly discrepant results on $\Omega$, 
but it is not too difficult to speculate their origins.
On the other hand, the current `low $\Omega$' means the values  
that vary almost by a factor of three and effort is needed to
make these converge.

\begin{figure}
\begin{center}
\centerline{{\epsfxsize=13cm\epsfbox{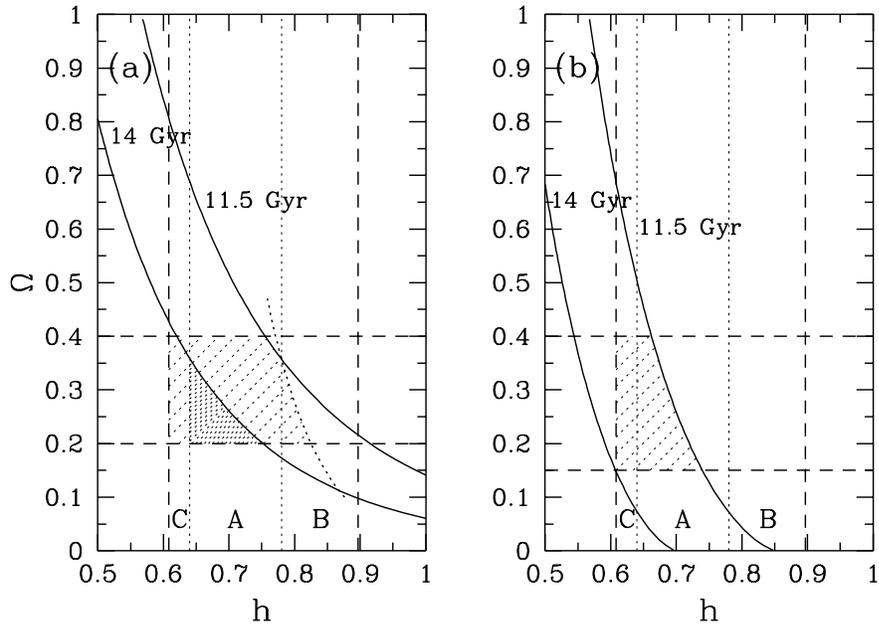}}}
\vspace{-4cm}
\caption{Consistent parameter ranges in the $H_0-\Omega$ space for 
(a) a flat universe and (b) an open universe. $A$ is the range of
the Hubble constant when $(m-M)_{\rm LMC}=18.5$. $B$ or $C$ is
allowed only when the LMC distance is shorter by 0.3 mag, or longer
by 0.1 mag. Note in panel (a) that most of the range of 
$B$ is forbidden by the compatibility of age and $H_0$ that 
are simultaneously driven by the RR Lyr calibration (section 2.7). 
Also note that the age range between $\approx$11.5 Gyr and $\approx$14 Gyr
is possible only with the interpretation that globular cluster 
formation is coevel (section 2.4). The most naturally-looking parameter
region is given a thick shade.} 
\end{center}	
\vspace{-1cm}
\end{figure}

The cosmological constant has been an anathema over many years because
of our ignorance of any mechanism that could 
give rise to a very small vacuum energy
of (3 meV)$^4$, and neither can we understand a zero cosmological constant.
In mid-nineties the atmosphere was changing in favour for a non-zero
$\Lambda$. The prime motivation was the Hubble constant$-$age problem,
but the introduction of a non-zero $\Lambda$ was helpful in many respects.
One theoretical motivation was
to satisfy flatness which is expected in inflationary scenarios (Peebles 1984).
Ostriker \& Steinhardt (1995) proclaimed a
`cosmic concordance' with a flat universe mildly dominated by $\Lambda$. 
By 1997, only one observation contradicted with the presence 
of a moderate value of $\Lambda$;
this was 
the SNeIa Hubble diagram presented by the
{\it Supernova Cosmology Project} (Perlmutter et al. 1997); see Fukugita 1997.
In the next two years the situation changed. Two groups 
analysing SNeIa Hubble diagram, including 
the {\it Supernova Cosmology Project},
now claim a low $\Omega$ and a positive $\Lambda$. On the other hand, the
Hubble constant$-$age problem became less severe due to our cognition
of larger uncertainties, especially in the age estimate.
The indications from SNeIa Hubble diagram are very interesting and 
important, but the
conclusions are susceptible to small systematic effects.
They should be taken with caution. 
We should perhaps wait for small-scale CBR anisotropy 
observations to confirm a nearly flat universe before concluding 
the presence of $\Lambda$.

In these lectures we have not considered classical tests, number counts,
angular-size redshift relations, and magnitude-redshift relations of
galaxies  (Sandage 1961; 1988), in those testing
for $\Omega$ and $\Lambda$. Unlike clusters or large scale structure,
where no physics other than gravity plays a role, the evolution of
galaxies is compounded by rich physics.
Unless we understand their astrophysics, these objects cannot be used as 
testing candles. It has been known that galaxy number counts is 
understood more naturally with a low matter density universe  under the
assumption that the number of galaxies are conserved, but it is
possible to predict the correct counts with an $\Omega=1$ model where galaxies
form through hierarchical merging, by tuning parameters that control
physics (Cole et al. 1994; Kauffmann et al. 1994). It
is important to work out whether the model works for any cosmological
parameters or it works only for a restricted parameter range. This does not
help much to extract the cosmological parameters, but it can falsify
the model itself.

We have seen impressive progress in the determination of the Hubble 
constant. The old discrepancy is basically solved. On the other hand,
a new uncertainty emerged in more local distance scales. The most
pressing issue is to settle the value of the distance to LMC. There are
also a few issues to be worked out should one try 
to determine $H_0$
to an accuracy of a 10\% error or less. 
They include understanding of metallicity effects and interstellar
extinction. The future effort will give more weight to geometric or
semi-geometric methods. From the view point of observations the work
will go to infra-red colour bands to minimise these problems.

In conclusion, I present in Figure 6 allowed ranges of $H_0$ and
$\Omega$ (and $\lambda$) for the case of (a) flat and
(b) open universes. With the flat case we cut the lower limit of $\Omega$
at 0.2 due to a strong constraint from lensing.
An ample amount of parameter space
is allowed for a flat universe. 
A high value of $H_0>82$, which would be driven only by a short
LMC distance, is excluded by consistency with the age of globular 
clusters as noted earlier. Therefore, we are led to the range
$H_0\simeq60-82$ from the consistency conditions.
For an open universe the coeval-formation
interpretation is compelling for globular clusters, or else no region
is allowed. The allowed $H_0$ is limited to $60-70$.
No solution is available if LMC takes a short distance. 

\vskip4mm
I would like to thank Rob Crittenden for his careful reading 
and many useful suggestions on this manuscript. 
This work is supported in part by Grant in Aid of
the Ministry of Education in Tokyo and Raymond and Beverly Sackler
Fellowship in Princeton.

\end{document}